# Spin-glass-like freezing of inner and outer surface layers in hollow γ-Fe$_2$O$_3$ nanoparticles

Hafsa Khurshid[1,*], Paula Lampen-Kelley[1], Òscar Iglesias[2], Javier Alonso[1,3], Manh-Huong Phan[1,*], Cheng-Jun Sun[4], Marie-Louise Saboungi[3,5], and Hariharan Srikanth[1,*]

[1] *Department of Physics, University of South Florida, Tampa, FL 33620*

[2] *Departament de Física Fonamental and Institut de Nanociència i Nanotecnologia (In$^2$UB)*
*Universitat de Barcelona, Av. Diagonal 647, 08028 Barcelona, Spain*

[3] *BCMaterials Edificio No. 500, Parque Tecnológico de Vizcaya, Derio, Spain 48160*

[4] *Advanced Photon Source, Argonne National Laboratory, Argonne, Illinois 60439*

[5] *IMPMC-Université Pierre et Marie Curie F-75252 Paris- Cedex 5 & Université d'Orléans*

**Disorder among surface spins is a dominant factor in the magnetic response of ultrafine magnetic particle systems. In this work, we examine time-dependent magnetization in high-quality, monodisperse hollow maghemite nanoparticles (NPs) with a 14.8±0.5 nm outer diameter and enhanced surface-to-volume ratio. The nanoparticle ensemble exhibits spin-glass-like signatures in dc magnetic aging and memory protocols and ac magnetic susceptibility. The dynamics of the system slow near 50 K, and become frozen on experimental time scales below 20 K. Remanence curves indicate the development of magnetic irreversibility concurrent with the freezing of the spin dynamics. A strong exchange-bias effect and its training behavior point to highly frustrated surface spins that rearrange much more slowly than interior spins with bulk coordination. Monte Carlo simulations of a hollow particle reproducing the experimental morphology corroborate strongly disordered surface layers with complex energy landscapes that underlie both glass-like dynamics and magnetic irreversibility. Calculated hysteresis loops reveal that magnetic behavior is not identical at the inner and outer surfaces, with spins at the outer surface layer of the 15 nm hollow particles exhibiting a higher degree of frustration. Our combined experimental and simulated results shed light on the origin of spin-glass-like phenomena and the important role played by the surface spins in magnetic hollow nanostructures.**

Recent advances in nano-fabrication techniques have allowed a rapidly growing body of research to take shape into the physical properties of unconventional nanostructures, including hollow architectures that exhibit very large surface-to-volume ratios, low density, and inner void space[1,2]. These versatile structures have found applications in fields ranging from energy (e.g. Li ion batteries to targeted drug delivery and diagnostics, gas sensing, and catalysis[3-6]. In technologies based on conventional nanoparticles (NPs), iron and iron-oxide compounds have long been the predominant materials of choice due in part to relatively large magnetic moment, biocompatibility, and well-established synthetic methods that afford good reproducibility, narrow size distributions, and a high degree of control over shape and size. The single-particle and collective magnetic responses of ferrite ensembles have been highly scrutinized in solid spherical NPs over the past two decades[7]. However, reports of static and dynamic magnetic properties in their hollow counterparts remain rather rare[8-11]. Surface-driven effects in the hollow morphology are amplified by larger surface area and enhanced anisotropy compared to a similarly sized solid particle. Away from the surface, interfacing between components of the internal structure of the granular hollow shell contributes to the complexity of the system.

The magnetic response of an ensemble of NPs is a combination of surface effects, finite size effects (i.e. truncation of the magnetic correlation length), and collective behaviors due to interparticle dipolar and/or exchange interactions. The primary exchange mechanism in ferrite compounds is an antiferromagnetic super-exchange interaction between metal cations mediated by an intervening oxygen ion. Because of the indirect nature of the coupling, the super-exchange interaction is sensitive to modified bond lengths and angles at a surface, and variation in coordination of surface cations will produce a distribution of net exchange fields. Oxygen vacancies and electronic bonding of organic surfactants further reduce the effective coordination of surface cations via broken exchange bonds[12,13]. The result is well-documented disorder in a layer of spins extending ~ 0.6 nm from the particle surface[14]. The magnetic properties of the surface layer are distinct from those of the interior region, and include reduced magnetization[12], spin canting[15], and a strong increase in effective anisotropy[7,15-18].

The presence of a second inner surface in a hollow particle greatly reduces the number of spins with bulk coordination as compared to conventional solid particles. In addition, Kirkendall-voided hollow NPs are composed of grains, each with their own magnetocrystalline anisotropy direction. The effective size of the interacting magnetic particles is thus related to the shell thickness rather than the total diameter of the hollow particle, and the magnetic properties of hollow NPs depend sensitively on the total particle diameter $d$ and shell thickness. For example, the intrinsic exchange bias observed in hollow $\gamma$-Fe$_2$O$_3$ hollow particles with $d \sim 20$ nm becomes a minor loop effect when the diameter is reduced to 10 nm as the large fraction of highly frustrated surface spins prevents saturation of the magnetization in fields up to several tesla[8,10]. In a recent study[11], we found that for $d < 10$ nm magnetic relaxation in a hollow particle ensemble was best described by a non-interacting particle model, as the dominant role of disordered surface spins and severely reduced particle magnetization rendered the influence of dipolar interactions negligible in determining low-temperature magnetic behavior. When the size of the hollow particles is increased to 15 nm, a sufficient portion of spins have bulk coordination (and thus bulk-like magnetic moment) to give rise to collective processes via dipole-dipole interactions, and relaxation could be described by the Vogel-Fulcher model for interacting particles.

Time-dependent magnetization in particles below the single domain limit is usually modeled in terms of thermally-activated relaxation between two stable magnetization states[19], separated by a well-defined energy barrier. Interparticle interactions modify these energy barriers so that they are no longer independent, and a crossover occurs from single-particle blocking to collective freezing of particle moments into a superspin glass (SSG) phase as a function of increasing interaction strength[20]. This scenario does not account for surface spin disorder, which gives rise to a more complex energy landscape with irregularly distributed barriers and multiple minima. Thus, the many degrees of freedom spins near a surface layer can also drive a spin-glass-like freezing at low temperatures. Glassy behavior in ferrite NP ensembles has been attributed to a SSG phase [20-23], and/or freezing of disordered surface spins[14, 25-27]. These phenomena are difficult to distinguish experimentally[27], and it is likely that both affect the slow dynamics of real systems, with relative contributions varying according to particle size and concentration.

In this work, we explore the role of surface spins in the static and dynamic magnetic response of an ensemble of 15 nm $\gamma$-Fe$_2$O$_3$ hollow particles. The results of our experimental observations and Monte Carlo simulations demonstrate that the freezing of disordered spins at the inner and outer surfaces leads to the development of spin-glass-like behaviors including memory, remanence, and aging effects as well as an exchange bias phenomenon in which surface spins play the role of an irreversible magnetic phase.

## Results and Discussion

**Temperature dependence of dc and ac susceptibilities.** Representative TEM images of the hollow NPs along with the accompanying selected area electron diffraction pattern (SAED) are shown in Fig. 1. The lack of contrast in the center of the particles confirms their hollow morphology. High-resolution TEM (HRTEM) micrographs reveal that the shell of each hollow particle is composed of randomly oriented grains, clustered together to form a hollow sphere with an average shell thickness of 3.25 ± 0.24 nm. A histogram representing the total outer diameters of more than 300 particles yields a mean particle diameter of 14.8 ± 0.5 nm when fit to a Gaussian distribution function. A quantitative analysis of the different Fe oxide phases present in the hollow shells was performed by fitting the X-ray absorption near edge spectroscopy (XANES) spectrum to a linear combination of different iron oxide forms using the Athena software package[30]. An excellent fitting is obtained using $\gamma$-Fe$_2$O$_3$ and Fe$_3$O$_4$ in the atomic proportion 82(6):18(4), confirming that our hollow NPs are comprised mainly of $\gamma$-Fe$_2$O$_3$.

Figure 2a presents the variation of magnetization with temperature, measured under an applied field of 50 Oe in the zero-field-cooled (ZFC), field-cooled-warming (FCW), and field-cooled-cooling (FCC) protocols. The peak temperature of the ZFC magnetization curve occurs near 60 K. This peak is conventionally taken as the blocking temperature $T_B$ associated with superparamagnetic NPs. The temperature dependences of magnetization under the FC and ZFC protocols show several noteworthy features: (1) in contrast with the typical behavior of superparamagnetic ensembles, in which the ZFC and FC curves converge just above $T_B$, the irreversibility between the ZFC and FC curves persists even at 300 K, far above the blocking temperature; this behavior may be attributed to increased anisotropy in the system or strong interparticle interactions. (2) Instead of the expected monotonic increase with decreasing temperature[31-34], the FCW magnetization reaches a maximum at ~ 43 K and then decreases below this temperature. A drop in the FCW magnetization below a critical temperature is usually associated with the collective freezing of the system and can indicate the presence of a spin-glass like phase[24,35]. This scenario is supported by (3) a striking thermal hysteresis between the FCW and FCC magnetizations that emerges below 50 K. In this case thermal hysteresis arises from slow dynamic processes in the glassy region that prevent the system from equilibrating over experimental time scales.

The results of Fig. 2a indicate that a more detailed analysis of magnetization dynamics in the hollow particle ensemble is necessary to classify their collective behavior. To this end, low-amplitude ($H_{AC}$ = 10 Oe) ac susceptibility curves were measured between 10 K and 300 K. As the frequency $f$ of $H_{AC}$ was varied between 10 Hz and 10 kHz, the peak position $T_p$ in the real part of the susceptibility $\chi'$ ($T$) shifted to higher temperatures (Fig. 2b, inset). The peak shift per frequency decade is quantified in the phenomenological parameter $K = \Delta T_p/(T_p \Delta \log f)$. The calculated value of $K = 0.051$ for the hollow particles is outside the expected range for superparamagnetic compounds ($K = 0.100 – 0.130$). Instead, the value is in the range of many canonical spin glass systems ($K = 0.005 – 0.06$), supporting the idea of spin-glass-like behavior in the hollow NPs. The characterization of the collective freezing process through which the system crosses into the low-temperature glassy regime, was performed by fitting $T_p$ (*f*) to the Vogel-Fulcher model of relaxation for interacting particles with uniaxial anisotropy (Fig. 2b),

$$\tau = \tau_o \exp[E_a/k_B(T - T_0)] , \qquad (1)$$

where $E_a$ is the anisotropy energy barrier, $\tau_o$ is the relaxation time of each nanoparticle, and $T_0$ is a characteristic temperature that provides a qualitative measure of the interparticle interaction energy. Fitted parameter values of $E_a = 570$ K and $\tau_0 = 1.2 \times 10^{-10}$ s, indicate strong anisotropy with relaxation times on the slower end of the superparamagnetic relaxation range ($\tau \sim 10^{-9}$s – $10^{-13}$s). The proximity of the fitted value $T_0 = 48.17$ K to the cusp in the FCW curve and onset of thermal hysteresis suggests that it may be attributed to the development of collective freezing among the particles below 50 K [36].

**Memory and aging effects.** To confirm spin-glass like behavior, aging and memory experiments were performed under ZFC and FC protocols. The memory effect is an experimental signature of spin glass systems[37]. In the FC memory protocol, the sample was cooled in a magnetic field of 50 Oe with intermittent stops at 90 K, 75 K, 50 K, 35 K and 15 K. At each stop, the field was switched off for $10^4$ s, then returned to 50 Oe before cooling resumed. Figure 3a shows the cooling magnetization curve $M_{IS}$ with intermittent stops and the memory curve $M_{Mem}$ taken during subsequent continuous warming up in a 50 Oe field. In $M_{IS}$, the magnetization drops during the wait time as the magnetic moments equilibrate in zero field. The magnitude of magnetization that is recovered when the field is switched back on depends on how quickly the moments realign in response to the applied field. In case of a spin glass, the dynamics slow down critically as the freezing temperature is approached, hindering the recovery and leading to a large step in the $M_{IS}$ ($T$) curve[38]. While the magnetization in Fig. 3a recovers almost completely after the intermittent stops at 90 K and 75 K, the magnetization steps are severe at 50 K and 35 K, confirming a rapid onset of glassy dynamics. At 15 K the magnetization step is once again vanishing. In this temperature range, the absence of the step reflects near complete spin freezing such that the decay in the magnetization when the field is switched off is negligible. As can be seen in the $M_{Mem}$ curve, on warming up the magnetization exhibits a kink at every intermittent stop at or below 50 K as the system recovers the lower energy magnetic configuration that was imprinted through redistribution of energy barriers during the cooling process[39].

While FC memory effects can be observed in superparamagnetic systems as well as in spin glasses, aging and memory effects under the ZFC protocol are unique to spin glasses[39]. For the ZFC memory experiment, the system was cooled down to 35 K from room temperature in zero field, aged for $10^4$ s, then further cooled to 5 K. At 5 K, a small field (50 Oe) was applied and the magnetization $M_{Mem, ZFC}$ was measured as the system was warmed up. When comparing the memory curve with a reference ZFC curve $M_{Ref, ZFC}$ it can be seen that $\Delta M$ ($T$) = $M_{Mem, ZFC}$ $-M_{Ref, ZFC}$ reaches a minimum at the intermittent stop temperature of 35 K (Fig. 3b). Non-zero $\Delta M$, observed between ~5 K and 60 K, develops as the system relaxes towards more stable configurations during the imposed waiting period, as described in both the droplet model and hierarchal energy model of spin glasses. In the droplet model, spin glass excitations form compact domains whose volume increases with time because of the non-equilibrium nature of the spin dynamics[30,41]. The volume of a droplet grows when the system is left unperturbed at a constant temperature during an intermittent stop and wait, with a simultaneous growth in its associated energy barrier that is frozen in on further cooling and retrieved on warming. The relatively lower energy barriers at the stop temperature for the reference ZFC curve result in greater thermally activated cluster flipping upon warming, increasing $M_{Ref, ZFC}$ over $M_{Mem, ZFC}$ near the stop temperature.

From previous results, the dynamics of the hollow nanoparticle system critically slow down below ~ 50 K, and are effectively frozen on the timescale of our measurement by 20-30 K. To confirm the crossover between slow and frozen relaxation dynamics in this temperature range, the system was cooled to 30 K in zero field. At 30 K a 50 Oe field was switched on and magnetization was recorded for 5000 s ($t_1$). In a second step, the system was cooled further to 20 K and magnetization was recorded for 7000 s ($t_2$). Finally, the system was re-heated to 30 K and the magnetization was measured for a period of 8000 s ($t_3$). The results are presented in Fig. 4. When the field is turned on initially at 30 K, the magnetic moments relax gradually toward the field direction and the magnetization grows logarithmically. Cooling to 20 K arrests this process and the magnetization remains constant between $t_2$ and $t_3$. When the temperature is returned to 30 K, the magnetization resumes its upward relaxation from its previous value. Moreover, the growth occurs along the same logarithmic path, as can be seen from the continuity in the joining of the curves starting from $t_1$ and from $t_3$ (inset, Fig. 4). This behavior is a manifestation of near complete freezing of the dynamics of the system between 20 K and 30 K.

**High-field experiments: Remanent magnetization and pinned spins.** The results presented above unambiguously establish the presence of frozen spins in the hollow nanoparticle ensemble at low temperatures. To determine how the fraction of total spins in the frozen state varies with temperature and applied magnetic field, $M(H)$ loops have been measured after ZFC and FC processes at various temperatures and under different cooling fields (Fig. 5a). A vertical displacement of the FC hysteresis loop favoring the cooling field direction is proportional to the number of frozen spins that cannot be reversed by the measurement field[42]. The net moment of these frozen spins can be quantified as

$$M_f = \frac{1}{2}\left[M(H^+) - M(H^-)\right], \quad (2)$$

where the cooling field defines the positive direction, and $M(H^+)$ and $M(H^-)$ are the maximum applied fields in the positive and negative directions, respectively. Figures 5b and 5c show the temperature dependence of the moment associated with the uncompensated frozen spins under cooling fields of 20 kOe and 50 kOe. It can be seen that in both cases, $M_f$ increases with decreasing temperature starting below ~50 K although $M_f$ is consistently lower for a 50 kOe field, indicating that the fraction of frozen spins is reduced in the presence of a large magnetic field. The second derivative of $M_f$ versus temperature ($d^2M_f/dT^2$) (Fig. 5c) peaks near 20 K, the freezing temperature identified in the previous section. This observation illustrates the alternative description of this critical temperature as the point of maximum rate of crossover of individual spins into the frozen spin state.

Irreversible magnetization was evaluated via isothermal remanent $M_{IRM}$ and thermo-remanent $M_{TRM}$ curves. In the TRM protocol, the system is cooled from room temperature to $T < T_f$ under a constant dc field, and $M_{TRM}$ is recorded immediately after switching the field off. In the IRM protocol, the system is zero-field cooled to $T < T_f$ where a dc field is switched on and stabilized. $M_{IRM}$ is then measured immediately after switching the field off. The remanent magnetization is an indication of the number of moments that are still oriented along the direction of applied field after the field is removed. Figure 6 shows the temperature dependence of IRM after applying and switching off fields of 10 kOe and 50 kOe, and TRM after cooling in and switching off a 50 kOe field. For $T > 50$ K, the remanence is close to zero. As the temperature is decreased below 50 K, $M_{IRM}$ and $M_{TRM}$ increase as the slowing dynamics of the system cause the magnetization to lag the applied field after it is switched off. Similarly strong increases in $M_{TRM}$ and $M_{IRM}$ have previously been attributed to enhancement in effective anisotropy at low temperature[43]. Since TRM is obtained under FC conditions, the magnetization at the moment the field is switched off will exceed that of the ZFC condition in which spins do not completely relax toward the field direction from the quenched zero-field disordered state, especially at low temperatures, and consequently $M_{TRM} > M_{IRM}$[44].

While $M_{TRM}$ increases monotonically with decreasing temperature, $M_{IRM}$ – which gives information about the switchable moments at a particular temperature – shows a maximum at 34 K under 10 kOe and at 22 K under 50 kOe, then decreases as the temperature is lowered. The peak-like behavior of the $M_{IRM}$ (T) curves is related to the freezing of spins into a random orientation during the ZFC process. While slow dynamics is a prerequisite for non-zero $M_{IRM}$, below the freezing temperature the magnetic fields required to rotate spins out of their quenched randomly oriented state become very large. The fraction of frozen spins grows with decreasing temperature (see Fig. 5), and so the magnetization acquired when fields of 10 kOe and 50 kOe are switched on in the IRM protocol comes primarily from the remaining reversible fraction of spins that closely follow the magnetic field. As no remanence is associated with the reversible spins and the frozen spins are not influenced by experimentally available magnetic fields and thus cannot contribute to the remanence magnetization, the IRM decreases with temperature. The drop in $M_{IRM}$ below 22 K under 50 kOe is in good agreement with the temperature range of spin freezing noted above. The $M_{IRM}$ peak shifts higher in temperature to 34 K when the field is reduced to 10 kOe, as greater thermal energy is required to activate dynamically frozen spins to allow their participation in remanent behavior when the barrier-lowering Zeeman energy is reduced.

The results of Fig. 5 and Fig. 6 give a clear indication of distinct reversible and irreversible contributions to the low-temperature magnetization in the hollow nanoparticle ensemble. As discussed above, the surface layer of a nanoparticle is a well-documented source of spin disorder and frustration due to defects, dangling bonds, and uncompensated spins. Previous studies have shown that both inner and outer surface spins in hollow NPs contribute to the enhancement of effective anisotropy and exchange bias effect[10], in contrast to solid NPs in which disorder is introduced from the outer surface only. The irreversible and reversible magnetization components indicated by our results can thus reasonably be assigned to the surface and interior spins in the shell of the hollow particles, respectively. The existence of an interface between spatially distinct hard and soft magnetic phases is borne out by the strong exchange bias effect (horizontal displacement of the hysteresis loop) of Fig. 5a. The dynamics of the spins at the irreversible/reversible interface were examined through the training effect of the exchange bias field, $H_{EB} = 0.5(H^+ + H^-)$, where $H^+$ and $H^-$ are the coercive fields of the ascending and descending branches of the hysteresis loop, respectively. The system was cooled in a field of 50 kOe to 10 K, where the magnetic field was cycled a number of times. From Fig. 7 it can be seen that $H_{EB}$ decreases strongly with successive cycling, as spin rearrangement occurs at the interface. Our results are described well by a model considering frozen $f$ and rotatable $r$ components with different relaxation rates at the interface[45],

$$H_{EB}^n = H_{EB}^\infty + A_f \exp(-n/P_f) + A_r \exp(-n/P_r), \qquad (3)$$

where the $A$ parameters are weighting factors with the dimension of magnetic field, the $P$ parameters resemble a relaxation rate, and $H_{EB}^\infty$ is the stable exchange bias field as $n \to \infty$. From the fit we obtain $H_{EB}^\infty = 962 \pm 101$ Oe, $A_f = 20.7 \pm 1.2$ kOe, $A_r = 1980 \pm 100$ Oe, $P_f = 0.39 \pm 0.01$, and $P_r = 2.74 \pm 0.79$. The ratio $P_r/P_f = 7.02$ indicates that the reversible spins rearrange 7 times faster than the irreversible spins at 10 K. At 35 K the exchange bias field is greatly reduced, and its training is not described well by (3) or more conventional power law decay models, but rather decreases in an approximately linear fashion.

**Monte Carlo Simulations** The geometry of a hollow particle dictates that the reversible interior spins form two distinct interfaces with irreversible spins at the inner and outer surfaces. We have performed Monte Carlo (MC) simulations of the hysteresis loops of a hollow particle with the same dimensions as in the current study in order to isolate the magnetic properties of spins forming the inner and outer surface layers of the shell. The simulations are performed at the atomistic level considering classical spins placed at the nodes of the spinel lattice of $\gamma$-Fe$_2$O$_3$[8,46]. The polycrystalline nature of the hollow particles as seen in TEM images (Fig. 1) is incorporated by dividing the simulation volume into a series of crystallites with equal volumes and randomly distributed anisotropy axes. The influence of external magnetic field on the alignment of spins in the inner and outer surface layers of the shell has thus been simulated (see Fig. 8, for example). Details of the simulations can be found in [7].

Figure 9 presents the results of a series of simulated hysteresis loops for a particle with surface Néel anisotropy with constant $k_s = 30$ K, and uniaxial anisotropy for core spins $k_c = 0.01$ K (equal to the bulk value of $\gamma$-Fe$_2$O$_3$) calculated at low temperature ($T = 0.1$ K) with a field step $dh = 5$. In Fig. 9a, magnetic hysteresis loops representing the total magnetization of the particle were calculated using different numbers of MC steps (MCS) to average the magnetization at every field in order to observe the influence of the field sweep rate and the relaxation dynamics of surface spins. Consistent with experimental observations, the loops have elongated shapes, high closure fields and linear high-field susceptibility indicative of a system with a high degree of disorder or frustration. In these hollow particles, frustration arises due to a large fraction of surface spins and random orientation of anisotropy axes of the crystallites. The resulting spin-glass-like state, observed experimentally at low temperature, is also reflected in the dependence of the coercive fields and remanent magnetizations on the field sweep rate – controlled in the simulation by the number of MCS. These quantities present a marked short-time dynamic evolution that stabilizes once surface spins attain their frozen state. The negative horizontal loop displacements and the difference between positive and negative remanent magnetization (that persists even at low cooling rates) are a signature of pinned frozen spins that cannot be equally reversed by a positive or negative field due to the complex energy landscape generated by disorder.

Figure 9b shows the contributions of the spins at the inner (long dashes) and outer (short dashes) surfaces of the hollow particle to the total hysteresis loop with MCS = 1000. It can be seen that inner surface spins are more easily magnetized than spins at the outer surface, and their associated hysteresis loops show a reduced vertical shift. These observations provide a clear indication of a change in dynamic properties of spins at the inner and outer surfaces, a consequence of a different range of effective energy barriers governing their relaxation. The differences in the response to a magnetic field at the two surfaces are illustrated in Fig. 8(a-d), where a slice of the particle along a plane perpendicular to the applied field axis shows the surface spins (interior spins are not depicted for clarity), colored according to their orientations as described in the legend. It can be seen that inner surface spins have tones closer to dark blue while outer spins closer to dark red both for high fields (Fig. 8a, h = 100) and low fields close to the coercive field (Fig. 8c, h = -25), indicating different degrees of disorder of spins in these two regions during the whole magnetization reversal process.

Finally, Fig. 9c and Fig. 9d compare the contribution of all surface spins and interior spins (those having bulk coordination) for the hollow nanoparticle (continuous black lines) to that corresponding to a solid particle with the same diameter (dashed red lines). The global shape of the hysteresis loops of surface spins is qualitatively similar in both cases. However, the reversal behavior of the interior spins is clearly influenced by the existence of additional surface spins at the inner surface of the hollow nanoparticle. As can be seen in Fig. 9d, the shift in the core hysteresis loop is increased for the hollow particle (this is also observed for the total contribution, not shown). Itsmore elongated shape together with reduced remanence and increased coercive field is a consequence of the increased disorder induced by the larger number of surface spins in the hollow case. As can be seen in Fig.9c, the coercive field and loop shifts for the surface contribution are almost identical for both morphologies while core contributions are different, which indicates that the additional inner surface spins in the hollows are responsible for the differences in the magnetic behavior with respect to the solid particle. Finally, notice also that a pinch of the loops around zero field values can be observed in the hollow particle although less pronounced than in the experimental results of Figs. 5 and 7. The fact that it is not visible for the solid particle, indicate that it may originate from inner surface spins.

In summary, we have investigated the static and dynamic magnetic properties of 15 nm $\gamma$-$Fe_2O_3$ hollow particles. The nanoparticle ensemble exhibits slow dynamics and aging behaviors below ~ 50 K. AC susceptibility and FC/ZFC memory and aging experiments show spin-glass-like characteristics in the system with the onset of a frozen spin state when the temperature is reduced below 20 K. Quantification of the magnetization associated with the frozen moments reveal a rapid growth in $M_f$ near 20 K, confirming a strong increase in the fraction of frozen spins at this temperature.

High-field FC and ZFC remanence curves indicate the coexistence of reversible and irreversible components of magnetization. The temperature at which a peak is observed in $M_{IRM}(T)$ (between 34 K and 22 K) suggests that the irreversible magnetic contribution derives from the previously identified dynamically frozen spins. A strong exchange-bias effect and its training behavior point to interfaces between reversible interior spins with bulk coordination and highly frustrated surface spins that rearrange up to 7 times more slowly. Atomistic Monte Carlo simulations confirm strongly disordered surface layers in the hollow particle morphology, with complex energy landscapes that underlie both glass-like dynamics and magnetic irreversibility. Calculated hysteresis loops show that inner surface spins are more easily magnetized than those located at the outer surface and produce a hysteresis loop with a smaller vertical shift, indicating that dynamic behaviors are not equivalent at the two surfaces. A stronger contribution to the experimentally observed spin freezing and related properties can thus be expected from the outer surface layer of 15 nm hollow particles.

## Methods

**Sample Preparation.** Hollow NPs were synthesized by oxidizing core/shell Fe/$\gamma$Fe$_2$O$_3$ NPs via the Kirkendall effect. Details of the synthesis procedure have been reported elsewhere[29]. In brief, iron NPs were chemically synthesized by the thermal decomposition of iron pentacarbonyl. A controlled oxidation of iron NPs facilitates the formation of a thin oxide layer near the particle surface giving rise to a core/shell type morphology with a metallic Fe core and iron oxide shell. In a subsequent step, the core/shell NPs were oxidized further at 180 °C in the presence of the high purity O$_2$. An imbalance in the diffusion rates of iron outwards from the core and oxygen inwards from the shell leads to the formation of voids between core and shell, which eventually coalesce to form a single cavity in the center of the particle. The final size of the resulting hollow particles is determined by the size of the Fe particle produced in the thermal decomposition step, which can be tuned by varying the injection temperature of iron pentacarbonyl[29]. For this study, hollow iron oxide NPs with an average diameter of ~15 nm were synthesized.

**Properties Characterization.** The structure, morphology, and composition of the products were characterized using an FEI Morgagni transmission electron microscope (TEM) and Fe K-edge XANES. XANES experiments were performed at the 20-BM-B beamline at the Advanced Photon Source (APS) at Argonne National Laboratory. A Si (111) monochromator crystal was used with energy resolution of about $\delta E/E=1.4 \times 10^{-4}$. The spectra were obtained in transmission mode, with a minimum step of 0.3 eV. The energy edge was carefully calibrated by simultaneously measuring an iron foil. To perform magnetic measurements, the NPs in solution were dried to a powder and packed firmly into a gelatin capsule to avoid physical motion of particles relative to one another. Magnetic measurements were carried out with a commercial Physical Property Measurement System (PPMS) from Quantum Design with vibrating sample magnetometer (VSM) and AC measurement system (ACMS) attachments.

# Acknowledgments

Research at the University of South Florida (NPs synthesis, structural and magnetic studies) was supported by the U.S. Department of Energy, Office of Basic Energy Sciences, Division of Materials Sciences and Engineering under Award No. DE-FG02-07ER46438. Research at the University of Barcelona (Magnetic simulation) was supported by Spanish MINECO (MAT2012-33037), Catalan DURSI (2009SGR856 and 2014SGR220), and European Union FEDER funds (Una manera de hacer Europa). Sector 20 facilities at the Advanced Photon Source, and research at these facilities, are supported by the US Department of Energy - Basic Energy Sciences, the

Canadian Light Source and its funding partners, the University of Washington, and the Advanced Photon Source. Use of the Advanced Photon Source, an Office of Science User Facility operated for the U.S. Department of Energy (DOE) Office of Science by Argonne National Laboratory, was supported by the U.S. DOE under Contract No. DE-AC02-06CH11357. Javier Alonso acknowledges the financial support provided through a postdoctoral fellowship from Basque Government. MLS acknowledges financial support from the CNRS, Paris, France and. We thank Tao Li, Bachir Aoun and Yang Ren for their help with the synchrotron measurements.

## Author contributions

H.K., P.L.K., O.I. and M.H.P. designed the study. The sample synthesis was performed by H.K. Structural and magnetic characterization and analysis were performed by H.K. and P.L.K. M.L.S. and C.J.S. performed XANES spectrum experiments. J.A. performed XANES linear combination analysis. O.I. performed Monte Carlo simulations. All authors discussed the results. H.K., P.L., O.I., M.H.P, and H.S. wrote the manuscript. H.S. led the research project.

## Additional information

**Competing financial interests:** The authors declare no competing financial interests.

**Figure Captions**

**Figure 1:** (a) TEM image of 15 nm hollow particles. Inset: SAED pattern indexed to fcc-structured iron-oxide. (b) High-resolution TEM image with crystallite boundaries marked by dashed lines. (c) Experimental XANES spectrum of hollow nanoparticle sample fitted with a linear combination of maghemite and magnetite contributions. The spectra for magnetite and maghemite are the weighted components used during the fitting.

**Figure 2:** (a) Temperature dependence of magnetization under zero-field cooled (ZFC) field cooled-cooling (FCC) and field cooled-warming (FCW) protocols. Inset: Expanded view of $M_{FCW}$ below 50 K. (b) Measurement period ($\tau = f^{-1}$ where $f$ is the frequency of $H_{AC}$) vs. peak temperature of real part of ac susceptibility $\chi'(T)$. Solid line represents fit to the Vogel-Fulcher model. Inset: $\chi'(T)$ vs. $T$ at various frequencies.

**Figure 3:** (a) FC memory effect experiment: Intermittent-stop cooling magnetization $M_{IS}$ and continuous warming memory curve $M_{Mem}$. (b) ZFC memory effect experiment: Difference $\Delta M$ in continuous warming memory curve $M_{Mem,ZFC}$ after single-stop ZF cooling and reference ZFC curve $M_{Ref,ZFC}$.

**Figure 4:** Magnetic relaxation with H = 50 Oe after zero-field cooling to 30 K ($t_1$), reducing temperature to 20 K ($t_2$), and reheating to 30 K ($t_3$). Inset: 30 K relaxation data only ($t_2$ not shown).

**Figure 5:** (a) 5 K $M(H)$ loops after cooling in zero field (ZFC) and 50 kOe (FC). (b) Temperature dependence of vertical shift $M_f$ in FC hysteresis loops with cooling fields of 20 kOe and 50 kOe. (c) Second temperature derivative of $M_f(T)$ with cooling fields of 20 kOe and 50 kOe.

**Figure 6:** Isothermal remanent magnetization ($M_{IRM}$) and thermomagnetic remanent magnetization ($M_{TRM}$) measured immediately after switching off magnetic fields of 50 kOe or 10 kOe.

**Figure 7**: (a) Cycled hysteresis loops measured at 10 K after field cooling in 50 kOe. (b) $H_{EB}$ as a function of the loop index (n) extracted from individual hysteresis loops. Red line represents fit of experimental data to Eqn. 3. Inset: Cycled hysteresis loops measured at 35 K after field cooling in 50 kOe.

**Figure 8:** Snapshots of the outer and inner surface spin configurations subject to varying magnetic fields (a) h = 100 (the maximum positive applied field), (b) h = 0 (remanence at the upper branch), (c) h = -25 (near the negative coercive field), and (d) h = -100 (the maximum negative applied field). Spins have been colored with a gradient from dark-red/dark-blue (outer/inner surface) for spins along the field direction to yellow/green (outer/inner surface) for spins transverse to the field direction. Only a slice of the spin configurations of a hollow nanoparticle close to the central plane and perpendicular to the field direction is shown.

**Figure 9:** Low temperature ($T$ = 0.1 K) simulated hysteresis loops for a hollow particle with the same dimensions as in the experiments (shell thickness of 2.5 nm) and surface and core anisotropy constants $k_S$=30 K and $k_C$ = 0.01 K . (a) $M(H)$ curves for the total magnetization at different field sweep rates, calculated using 100, 500 and 1000 MCS to average the magnetization at every field. (b) Contributions of the spins at the inner (long dashed lines) and outer (short dashed lines) surfaces of the hollow particle to the total hysteresis loop with MCS = 1000. (c) Contribution of all surface spins to the total magnetization for the hollow NP (continuous black line) is compared to that of a solid NP with the same diameter (dashed red line). (d) The same as in panel (c) but for the interior spins contribution.

**Figure 1**

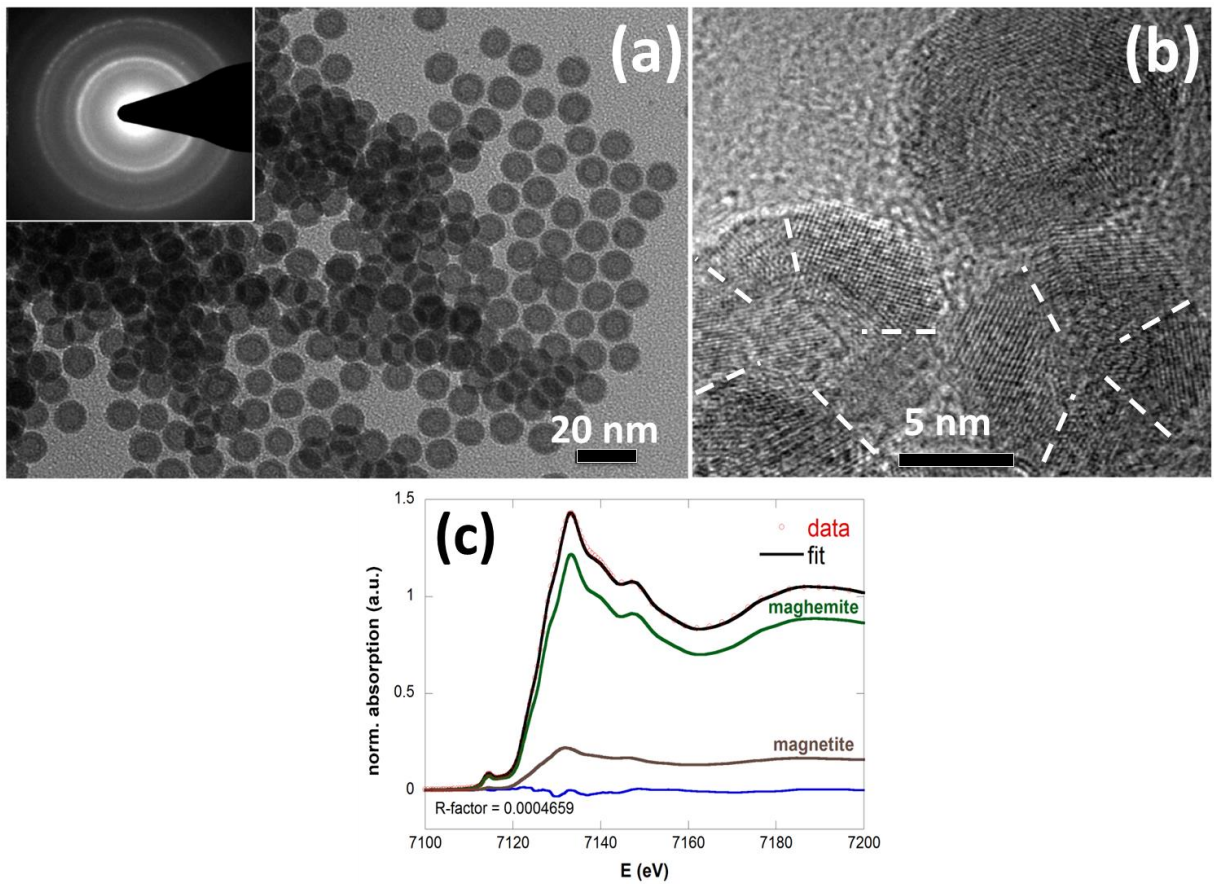

**Figure 2**

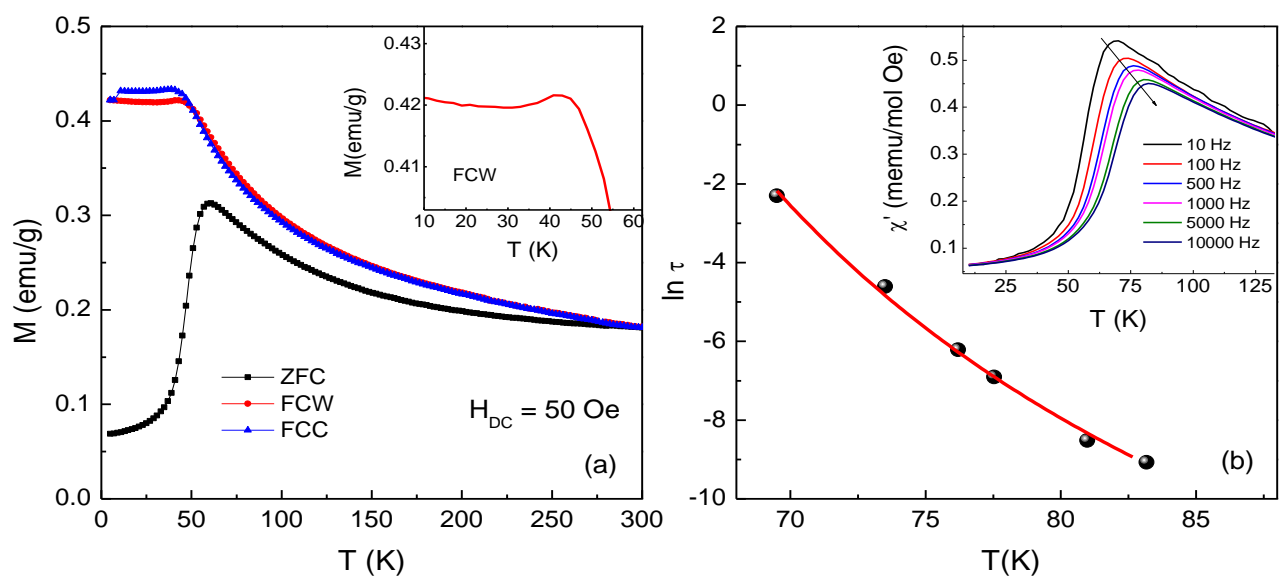

**Figure 3**

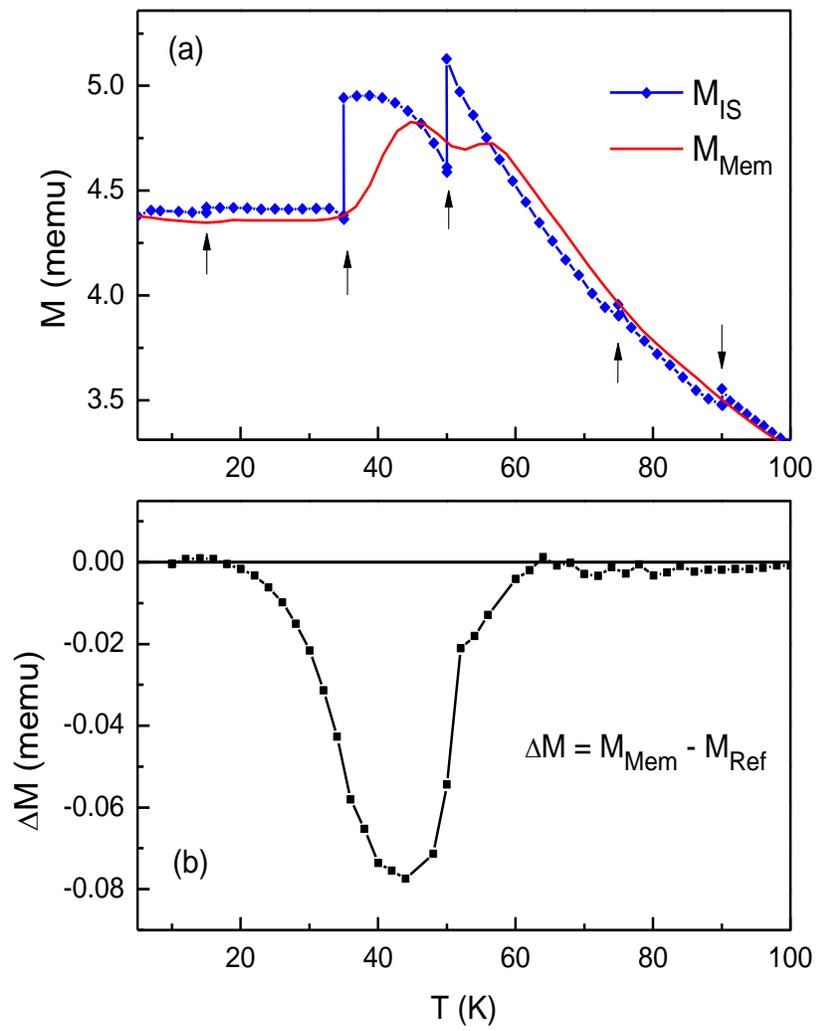

**Figure 4**

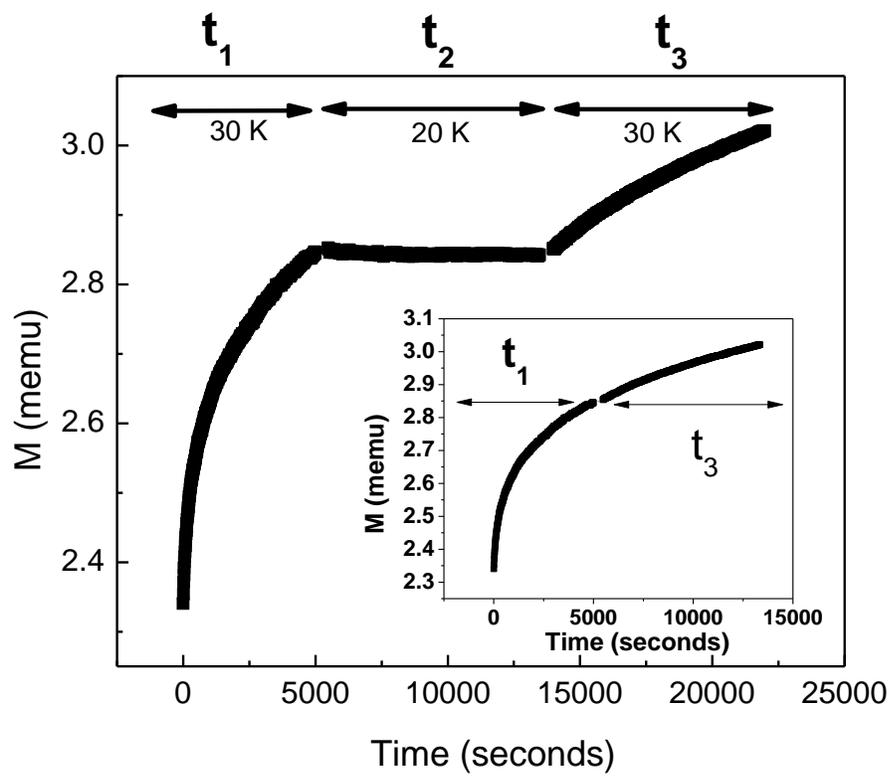

**Figure 5**

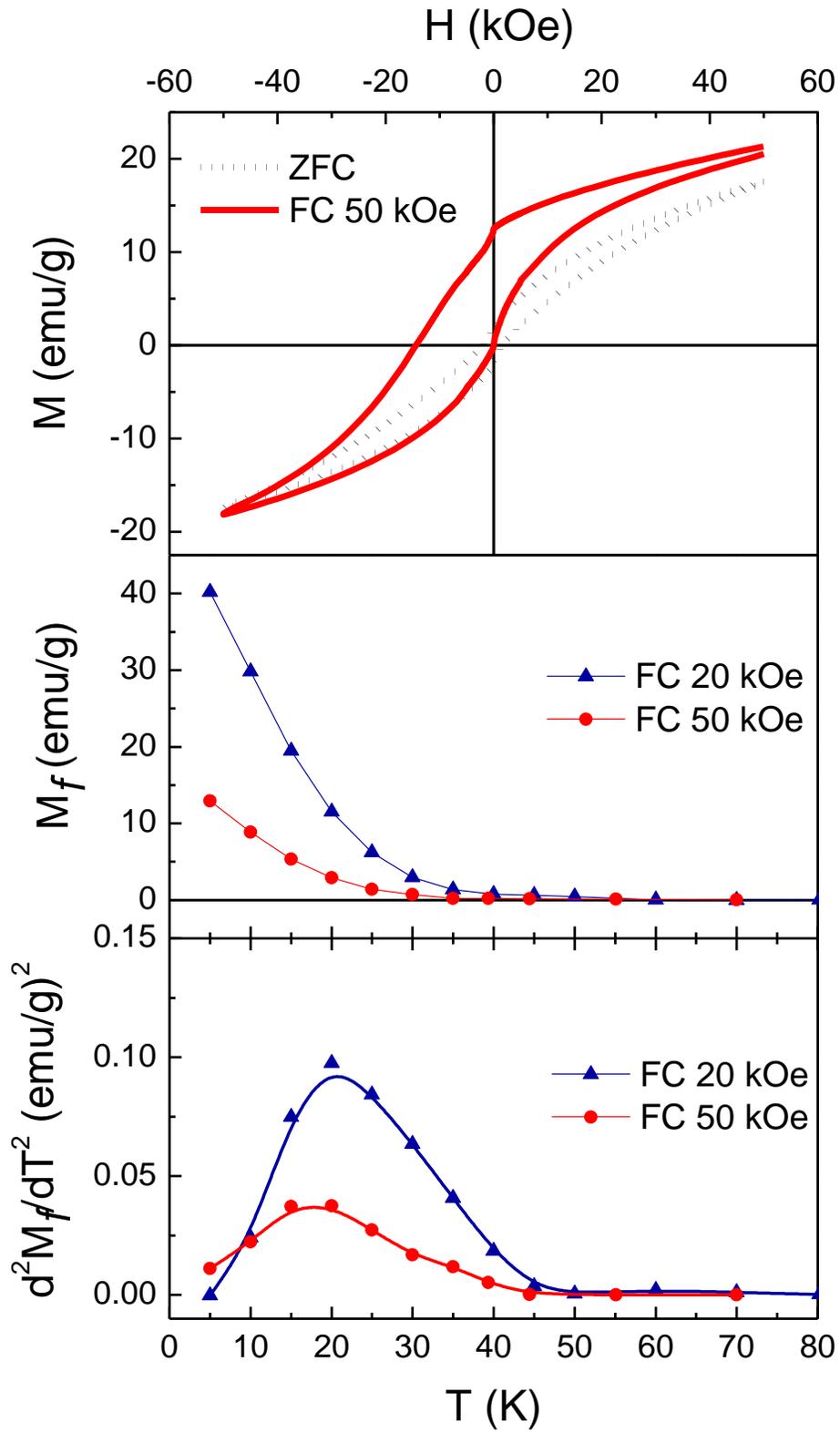

**Figure 6**

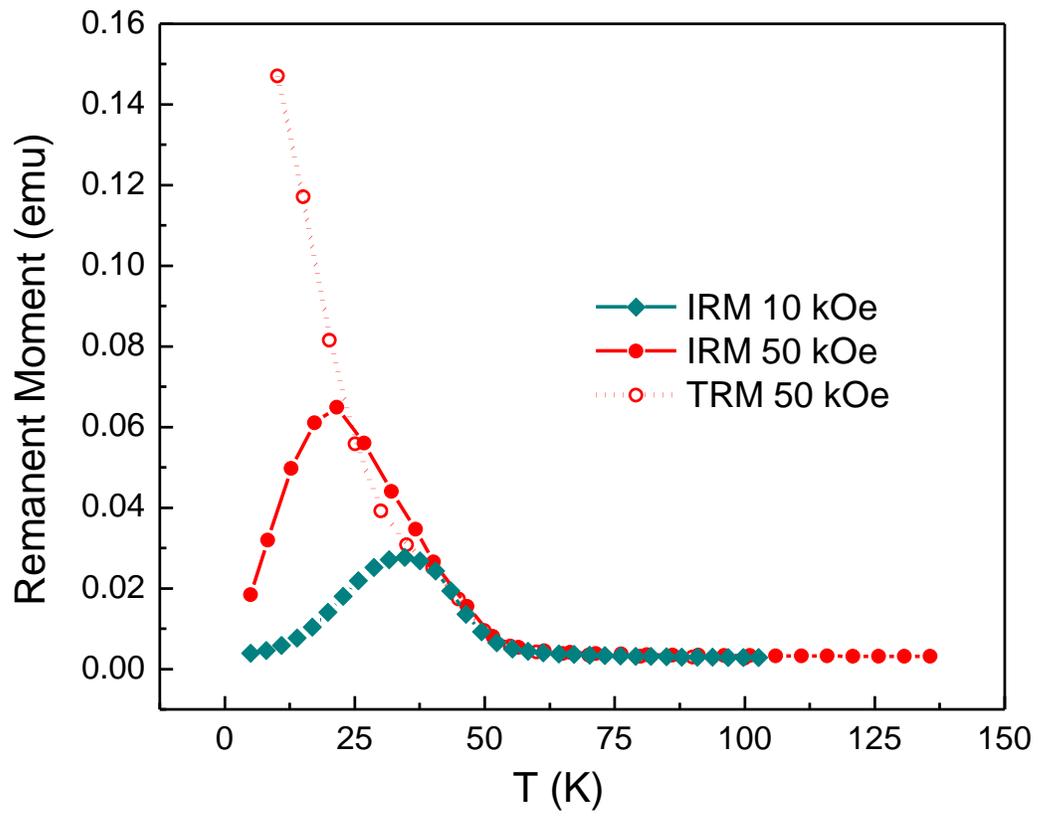

**Figure 7**

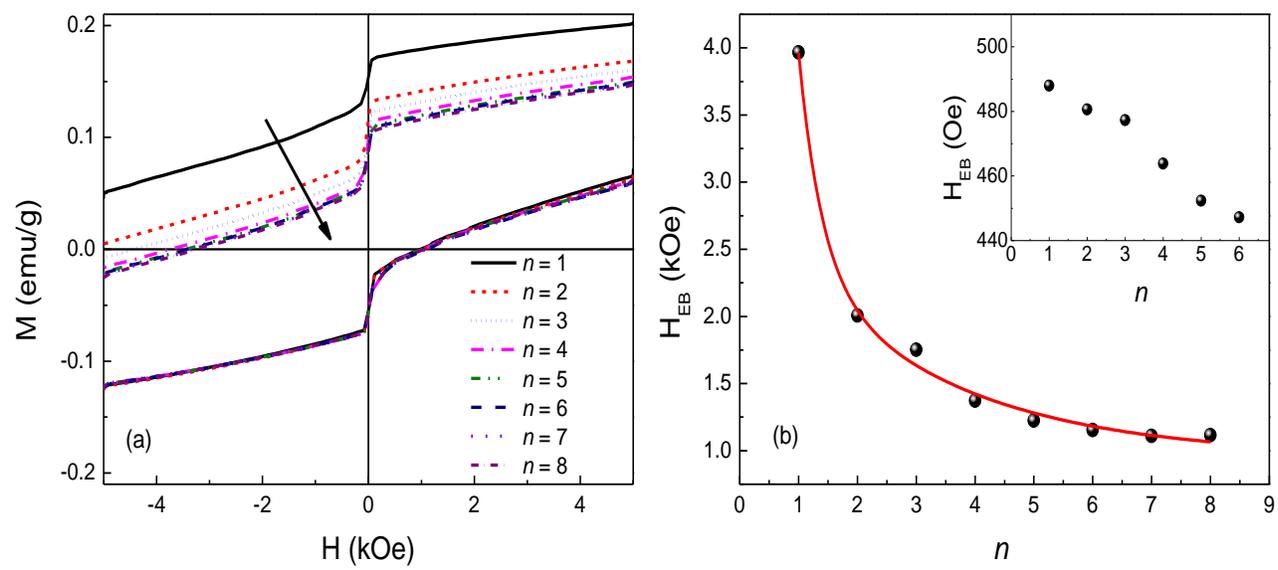

**Figure 8**

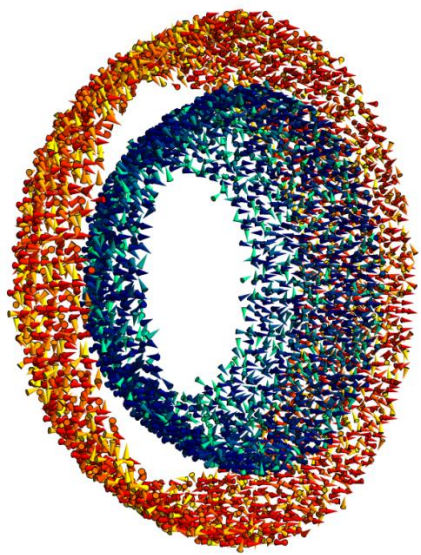

(a) h = 100

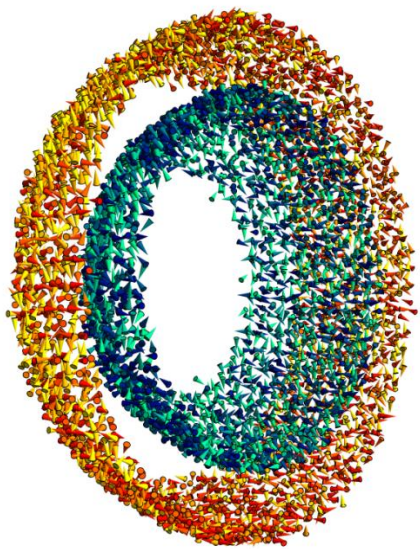

(b) h = 0

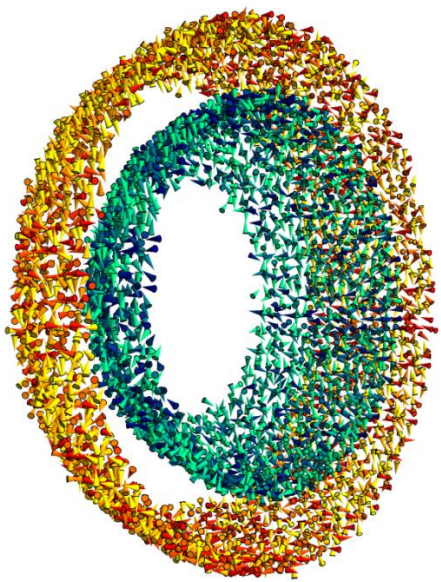

(c) h = -25

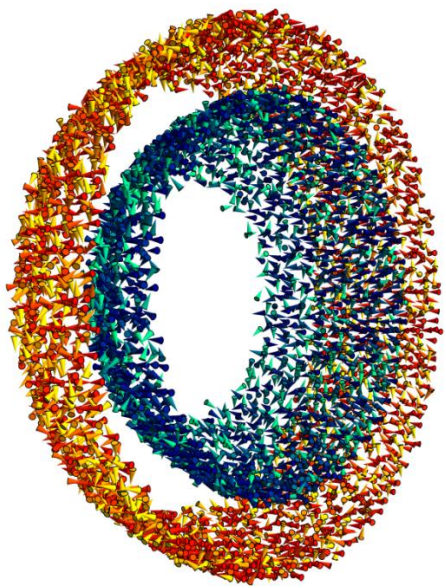

(d) h = -100

**Figure 9**

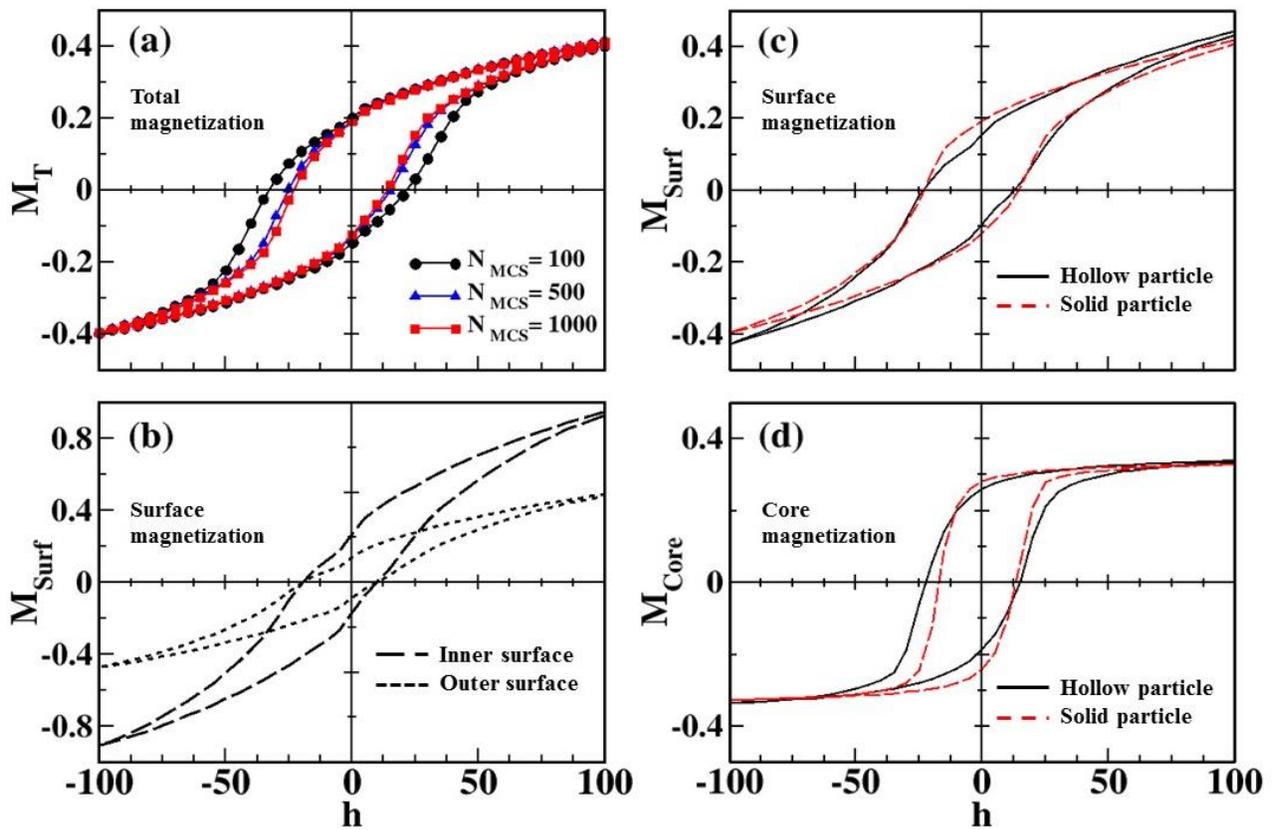